\documentclass[12pt]{article}
\usepackage{amsmath}
\usepackage{graphicx}
 \newcommand{\bea}{\begin{equation}}
 \newcommand{\eea}{\end{equation}}
 \newcommand{\ber}{\begin{eqnarray}}
 \newcommand{\eer}{\end{eqnarray}}
 \newcommand{\nn}{\nonumber}
\begin{document}

\title{BROWNIAN MOTION-PAST AND PRESENT}
\author{D.Chakraborty \footnote {e-mail:tpdc2@mahendra.iacs.res.in },
H.S.Samanta\footnote {e-mail:tphss@mahendra.iacs.res.in},
J.K.Bhattacharjee\footnote{e-mail:tpjkb@mahendra.iacs.res.in}\\
Department of Theoretical Physics\\
Indian Association for the Cultivation of Science\\
Jadavpur, Calcutta 700 032, India}

\maketitle
\begin{abstract}
We discuss Brownian motion from the more elementary viewpoint presented by Einstein in 1908. Later
developments and applications are briefly reviewed.
\end{abstract}
One of the celebrated papers of 1905 deals with a quantitative
study of the random motion.A couple of years later Einstein wrote 
a simplified account of the seminal work and we will present this 
version in this article. A direct test of the molecular kinetic theory was his obsession 
during his student days.The analysis of motion of suspended particles
in a liquid was an attempt to make observable predictions that when 
tested would establish,beyond doubt, the correctness of the molecular point
of view. \\

The analysis begins by considering a non-uniform solution in a container
with a semi-permeable membrane of thickness $\Delta x$ in the centre,as shown in Fig 1.
We need to consider the force balance in the region of width $\Delta x$. The 
pressure will be different on either side of the membrane because of the 
differing concentration. This will cause a force $\displaystyle{-\frac{\partial P}{\partial x} A \Delta x}$ 
to the left, where $ A $ is the cross-sectional area. Clearly, if the pressure decreases 
as $x$ increases, then the direction of the force will be to the right. If the local 
density gradient is negative, that is the concentration is higher to the left, 
then there will be a drift of the solute molecules to the right. If the drift speed $ v $
and the shear viscosity of the solvent is $\eta$, then there will be a resistive force
$ 6\pi \eta r v \textrm{ (Stokes Law)} $ on the solute molecule, if it is taken to
be a sphere of radius $r$. These are the only forces in the horizontal direction 
and a steady state would require that the net external force is zero.\\
\begin{figure}
\centering
\includegraphics[width=8cm,height=6cm]{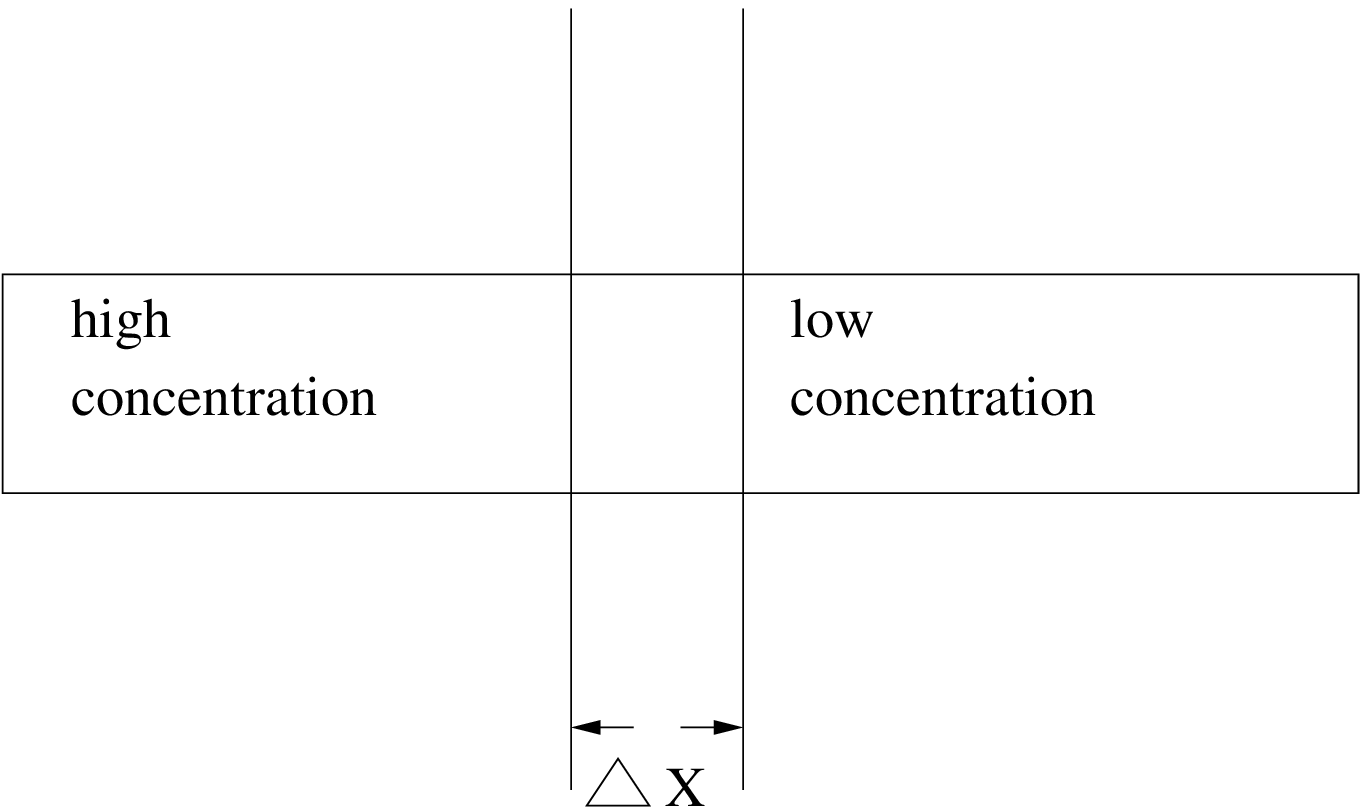}
\caption{}
\end{figure}

If $n(x)$ is the local density of solute molecules, then the total number in
the volume $ \displaystyle{A \Delta x} $ is $ {n(x) A \Delta x}$ and the total viscous force will 
be $ \displaystyle{6\pi \eta\, r\, v\, n(x) A \Delta x} $. The force balance then yields 
\bea
 6\pi \eta \,r\, v\, n(x) A \Delta x = -\frac{\partial P}{\partial x} A \Delta x \nn
\eea
or\\
\bea
\label{1}
v=-\frac{1}{6\pi \eta\, r\, n}\frac{\partial P}{\partial x}
\eea
The solution pressure can be written as 
\bea
\label{2}
P=n k T  
\eea
at temperature $T$ and thus 
\bea
\label{3}
v=-\frac{1}{6\pi \eta r n}\frac{\partial n}{\partial x}
\eea
The current of the solute molecules is the number of them crossing an unit area
per unit time and hence can be written as 
\bea
\label{4}
j=n v
\eea
From Eq.(\ref{3}), we find
\bea
\label{5}
j=-\frac{k_BT}{6\pi\, \eta\, r}\frac{\partial n}{\partial x}
\eea
The diffusion coefficient $D $ is defined by the relation 
\bea
\label{6}
j=-D\frac{\partial n}{\partial x}
\eea
and thus from Eq.(\ref{5}) and Eq.(\ref{6}) we have
\bea
\label{7}
D=\frac{k_BT}{6 \pi \eta r}
\eea

We now look at the statistical property of the current. In the container 
shown in Fig.(1), we consider any cross section (the solid line in Fig. 2) AB and
\begin{figure}
\centering
\includegraphics[width=8cm,height=6cm]{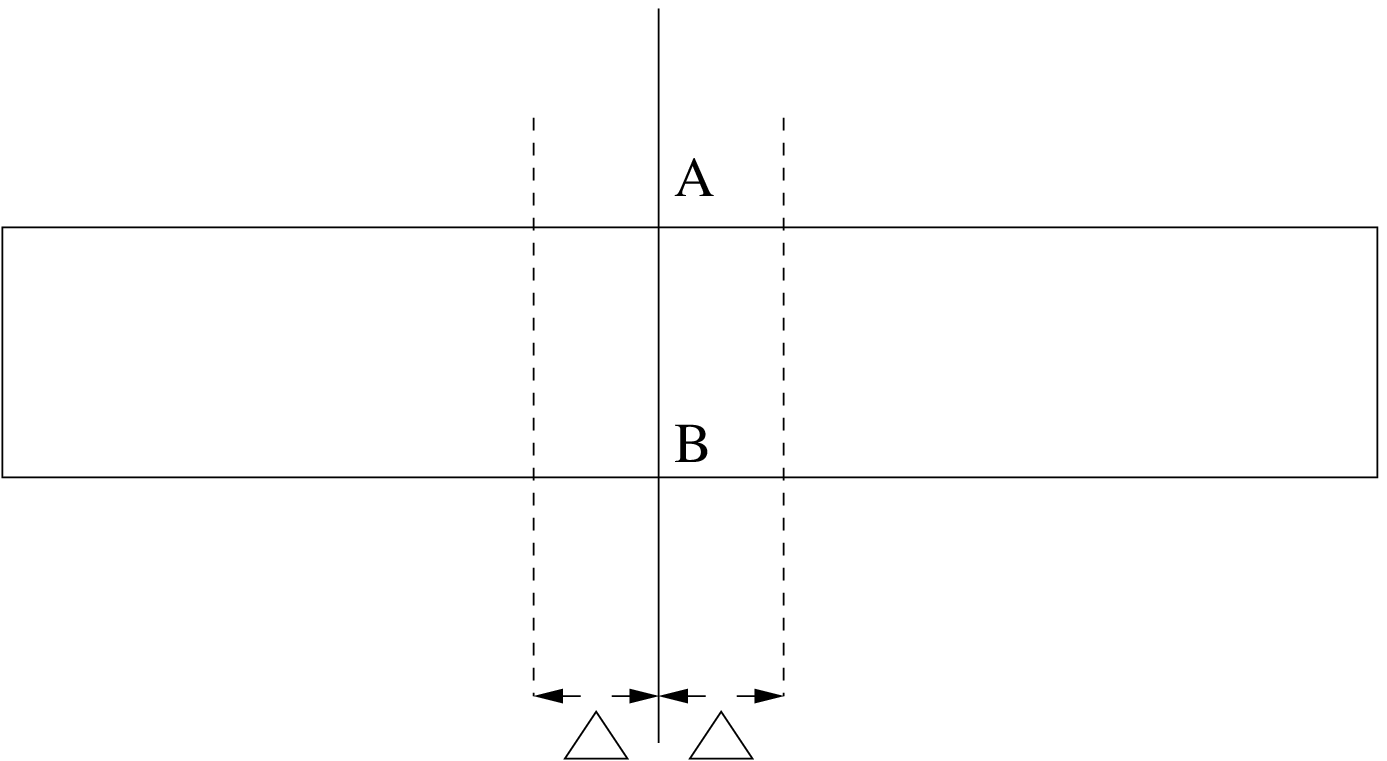}
\caption{}
\end{figure}
would like to calculate the current across the section due to the random hopping 
of the solute molecules. This random motion is produced by the random collisions
of the solute molecules with the molecules of the solvent. In a time $\tau$, 
we assume the root mean square displacement of a solute particle to $\triangle$.
Consequently, a current across the section AB will be set up by those solute 
particles on the left of it that are moving to the right and solute particles 
to the right of AB that are moving to the left. Since the probabilities of 
a left or right jump are equal, half the particles within a distance $\triangle$
to the left of AB will cross AB in time $\tau$ and half the particles 
within a distance $\triangle$ to the right of AB will cross it in time $\tau$.
If $n_l$ is the density of the particles to the left and $n_r$ the density of the 
particles to the right of AB, then the number $n$ of particle crossing an unit 
area of AB in time $\tau$ is given by
\ber
\label{8}
n &=& \frac{1}{2} (n_{l} \triangle - n_{r} \triangle) \nn \\
 &=& \frac{1}{2}\triangle (- \frac{d n}{d x} )
\eer
Hence, the number crossing unit area per unit time (which is the current) is
\bea
\label{9}
j=-\frac{1}{2} \frac{\triangle^{2}}{\tau} \frac{d n}{d x}  
\eea
Comparing with Eq.(6), we have the second important conclusion of Einstein 
\bea
\label{10}
 \triangle^2=2 D \tau
\eea
Between Eq.(\ref{7}) and Eq.(\ref{10}), one has enough predictive power to find the diameter
of suspended particles or Boltzmann's constant ${ k_B }.$

The verification of Einstein's predictions was primarily the work of Perrin and his
students. Early indications of the correctness of Eq.(\ref{7}) and Eq.(\ref{10}) came from 
the observations of Seddig, who took two photographs of an aqueous suspension of 
cinnabar on the same plate at an interval of 0.1 second and measured the distance
of corresponding images on the plate. He found that the distances at different temperatures were
inversely proportional to viscosity as predicted. Perrin and his students 
followed the movements of single particles of gamboge or mastic under a microscope
and recorded their positions at equidistant time intervals by means of an indicating
apparatus. Since the particle size was known, these observations yielded $k_B$ and since
$R$ was known, one could get $\displaystyle{N=\frac{R}{k_B}}$.\\
Perrin explicitly established that the suspended particles were in thermal equilibrium
with the solvent by studying the distribution of particles in a vertical column under the 
action of gravity. The total energy of a suspended particle at a height $z$ above the base
is $\displaystyle{\frac{p^2}{2m}+mgz}$ and hence the number of particles between $z$ and $z+dz$ is given by
\bea
\label{11}
n(z)=C A \int \frac{d^3 p}{h^3} \,exp(- \frac{p^2}{2mk_BT}) \, exp(- \frac{mgz}{k_BT})
\eea
where $C$ is a constant and $A$ is the area of the base of the container. 
The total number of particles is found from
\bea
\label{12}
N=C A \int \frac{d^3 p}{h^3} \, exp(- \frac{p^2}{2mk_BT}) \int \,exp(- \frac{mgz}{k_BT})
\eea
From Eq.(\ref{11}) and Eq.(\ref{12}) we have
\bea
\label{13}
\displaystyle{n(z)=N \frac{exp(-\frac{mgz}{k_BT})}{\int_{0}^{h} exp(-\frac{mgz}{k_BT})}=
N\frac{mg}{k_BT} \frac{exp(-\frac{mgz}{k_BT})}{\bigl[1- exp(-\frac{mgh}{k_BT})\bigr]}}
\eea

\begin{figure}
\centering
\includegraphics[width=5cm,height=8cm]{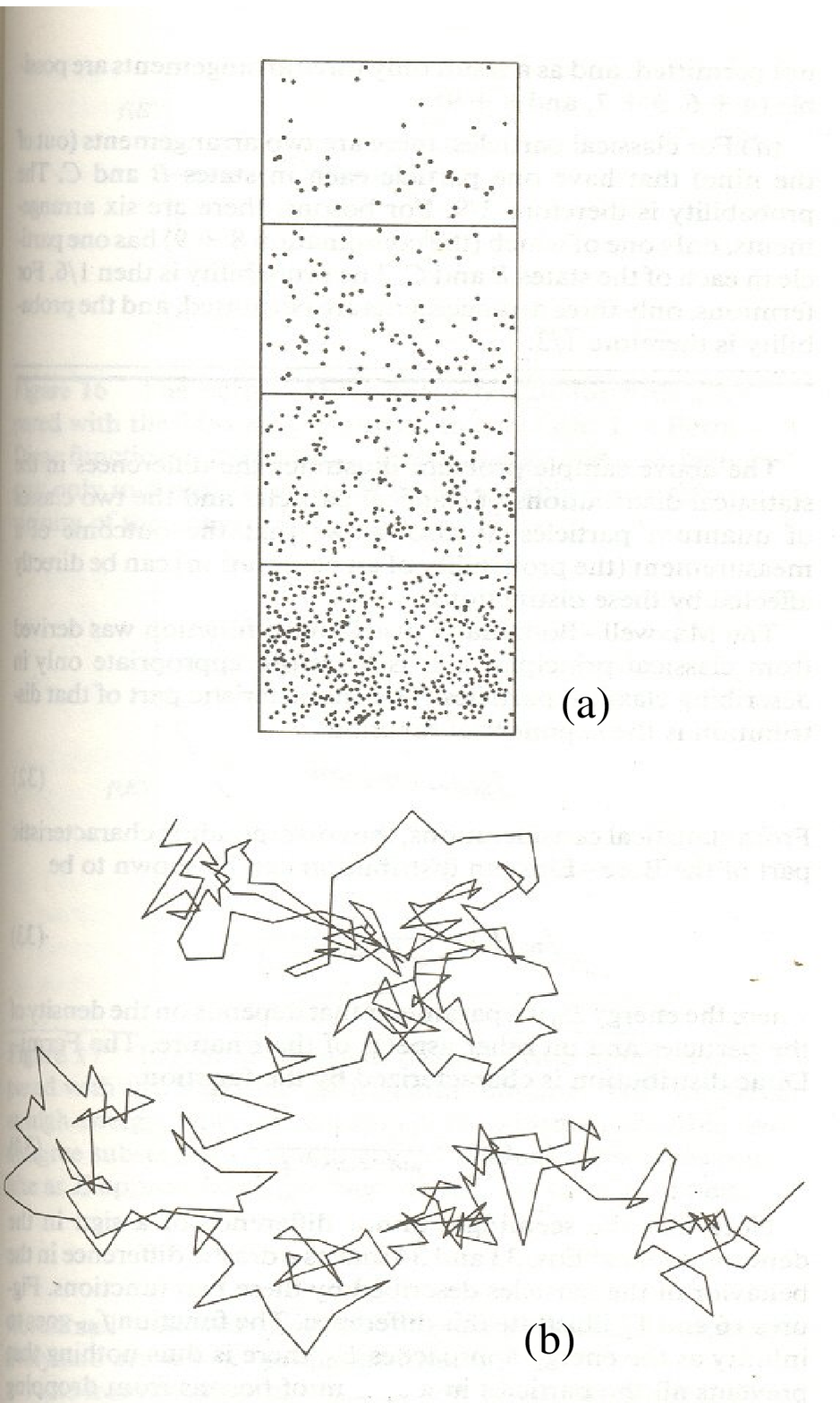}
\caption{}
\end{figure}
 
This distribution was confirmed in Perrin's experiment (Fig. 3a). The displacement 
of individual particles had the typical from shown in Fig 3b. The path of the 
particle is an example of a fractal, a curve for which any small section resembles 
the curve as a whole.

A different way of looking at the problem was derived by Paul Langevin, who was a 
friend of Einstein. Consider a molecule of mass $M$ colliding elastically with 
another molecule of mass $m$. If the velocities of $M$ before and
after the collision are $V$ and $V'$ respectively and those of $m$ are $v$ and $v'$ then
\bea
\label{14(a)}
M \vec V+m \vec v=M \vec V'+m \vec v'   
\eea
and
\bea
\label{14(b)}
\frac{1}{2} MV^2 + \frac{1}{2} mv^2 = \frac{1}{2} MV'^2 + \frac{1}{2} mv'^2
\eea
From the above equations we immediately find that
\bea
\label{15}
V'= \frac{M-m}{M+m} \vec V + \frac{2m \vec v}{M+m}
\eea
We now imagine that the molecule $M$ is being hit randomly by the molecules $m$ from all directions. 
In each collision, the change in momentum is
\bea
\label{16}
\Delta \vec P= M (\vec V-\vec V') = - \frac{2mM}{M+m} \vec V +\frac{2mM}{M+m} \vec v
\eea 
The force on $M$ due to the random collisions, is then clearly composed of two parts :\\
\\
 i) a resistive part which is proportional to $\vec V$. \\
\\
ii) a random part which is proportional to the velocity of the irregular motion of the 
molecule $m$. The average value of the force is zero and the mean square value is proportional to
$k_BT$, where $T$ is the temperature because the mean square velocity of the gas molecule 
is proportional to $k_BT$.

The equation of motion for the molecule can be written as 
\bea
\label{17}
\frac{d \vec V}{dt}= - \Gamma  \vec V + \vec f
\eea
where $ \Gamma $ is the relaxation rate and $\vec f$ is a random force with
\ber
\label{18(a)}
<f_i>& = & 0\\ 
\label{18(b)}
<f_i(t)f_j(t')> &=& 2 \sigma \delta (t-t') \delta_{ij}
\eer
The above arguments tell us that $\displaystyle{\sigma \propto k_BT}$ and if $M$ is a spherical solute molecule
moving through the fluid, then $ \displaystyle{\Gamma = \frac{6 \pi \eta r}{M}}$ where $r$ is the radius. 
The relation between $\sigma$ and diffusion constant $D$ is important for the establishment of 
thermal equilibrium.

The solution of Langevin equation can be written down as 
\bea
\label{19}
V_i(t)=e^{\Gamma t} \int_{0}^{t} e^{\Gamma t'} f_i(t') dt' + V_i(0)e^{-\Gamma t}
\eea
The mean square velocity is can be written as 
\ber
\label{20}
<V_i^2(t)>=<e^{-2 \Gamma t} \int_{0}^{t} e^{\Gamma t'} f_i(t')dt' \int_{0}^{t} e^{\Gamma t''} 
f_i(t'')dt''> +        \nn \\
 V_{i}^{2}(0)e^{-2 \Gamma t} + 2<V(0)e^{- \Gamma t} \int_{0}^{t} e^{\Gamma t'} f_i(t')dt'>
\eer
The averaging is over the random force $f(t)$ and using Eq.(\ref{18(a)}) and Eq.(\ref{18(b)}) we have
\ber
\label{21}
<V_i^2(t)>&=&2 \sigma e^{-2 \Gamma t} \int e^{2 \Gamma t'} dt' +V^2(0)e^{-2 \Gamma t} \nn \\
          &=& \frac{\sigma}{\Gamma} + (V_{i}^{2}(0) - \frac{\sigma}{\Gamma}) e^{- 2 \Gamma t}
\eer
If the solute molecules come into thermal equilibrium with the solvent molecules, then they too will
have a Maxwell Boltzmann velocity distribution in which case $<v_{i}^{2}>=\frac{k_BT}{M}$ and we see that 
$\frac{\sigma}{\Gamma}=\frac{k_BT}{M}$. The displacement $\triangle x_i$ in time t can be written as 
\ber
\label{22}
\triangle x_i & = & \int_{0}^{t} V_i(t')dt' \nn \\
             & = & \int_{0}^{t}dt' e^{-\Gamma t'}\int_{0}^{t'} dt_1 e^{\Gamma t_1} f_i(t_1) 
+\frac{V_i(0)}{\Gamma}(1-e^{-\Gamma t})
\eer
This leads to 
\ber
\label{23}
<(\triangle x_i)^2>=<\int_{0}^{t} dt' e^{-\Gamma t'} \int_{0}^{t'} e^{\Gamma t_1} f_i(t_1) dt_1
\int_{0}^{t} dt'' e^{-\Gamma t''} \int_{0}^{t''} e^{\Gamma t_2} f_i(t_2) dt_2>\\ \nn
+\frac{V_i^2(0)}{\Gamma^2} (1-e^{-\Gamma t})^2 + 
2 \frac{V_i(0)}{\Gamma} (1-e^{-\Gamma t})
<\int_{0}^{t} dt' e^{-\Gamma t'} \int_{0}^{t'} e^{\Gamma t_1} f_i(t_1) dt_1>
\eer
Using the moments of $f(t)$
\ber
\label{24}
<(\triangle x_i)^2>&=&2 \sigma \int_{0}^{t}dt' \int_0^t dt'' e^{-\Gamma (t'+t'')} 
\biggl[ \int_0^{t' } dt_1 \, exp(2 \Gamma t_1) 
\theta(t''-t') \nn \\
&+&\int_0^{t''} dt_1 \,exp(2 \Gamma t_1) \theta (t'-t'') \biggr]  
 + \frac{V_i^2(0)}{{\Gamma}^2}[1-exp(-\Gamma t)]^2 \nn\\
 &=&\frac{\sigma}{\Gamma}\biggl[ \int_0^t dt'\int_0^t dt''\, exp(-\Gamma (t''-t')) \theta(t'-t'') \nn \\
 &+& \int_0^t dt' \int_0^t dt'' exp(-\Gamma (t'-t'')) \theta(t'-t'')\biggr] 
 +\frac{V_i^2(0)}{{\Gamma}^2}[1-exp(-\Gamma t)]^2  \nn\\
&=&\frac{\sigma}{\Gamma}\biggl[ \int_0^t dt'\int_0^t dt''\,exp(-\Gamma (t''-t'))
+\int_0^t dt' \int_0^t dt'' exp(-\Gamma (t'-t''))\biggr] \nn\\
&+&\frac{V_i^2(0)}{{\Gamma}^2}[1-exp(-\Gamma t)]^2 \nn \\
&=&\frac{\sigma}{\Gamma^2}\biggl[\int_0^t dt'' [1-exp(-\Gamma t'')] +\int_0^t dt' [1-exp(-\Gamma t')]\biggr]
+\frac{V_i^2(0)}{{\Gamma}^2}(1-e^{-\Gamma t})^2 \nn \\
&=& \frac{2\sigma}{\Gamma^2} t + \frac{2 \sigma}{\Gamma^3} (1-e^{-\Gamma t}) 
+\frac{V_i^2(0)}{\Gamma^2}(1-e^{-\Gamma t})^2 \nn \\
& &
\eer
For extremely long times, the first term dominates and
\ber
\label{25}
<(\triangle x_i)^2>&=&\frac{2\sigma}{\Gamma^2} t \nn \\
&=&2 \frac{k_BT}{M} \frac{t}{\Gamma} \nn \\
&=& 2 \frac{k_BT}{6\pi \eta r}t \nn \\
&=& 2Dt 
\eer
where $\displaystyle{D=\frac{k_BT}{6 \pi \eta r}}$ is the Einstein relation as found in Eq.(\ref{7}) and Eq.(\ref{10}).
There is yet another way of approaching the problem. This is through the Fokker-Planck equation
which gives an equation for the dynamics of the probability distribution associated with the 
random process described by the Langevin description of Eq.(\ref{17}). If $P(x,t)$ is the probability 
distribution associated with the random variable $x(t)$ which satisfies the Langevin equation
\bea
\label{26}
\frac{dx}{dt}=-\Gamma \frac{\partial S}{\partial x} +f(t)
\eea
with the random force being Gaussian and having the correlation
\bea
\label{27}
<f(t_1)f(t_2)>= 2 D \delta(t_2-t_1)
\eea
Then the probability $P(x,t)$ of the variable $X(t)$ having a value $x$ at time $t$ satisfies
the evolution equation
\bea
\label{28}
\frac{\partial P}{\partial t} = \frac{\partial}{\partial x}(\Gamma \frac{\partial S}{\partial x} P)
+D \frac{\partial^2 P}{\partial^2 x}
\eea
In the case of a particle simply changing it's position because of random fluctuations we can drop the 
deterministic part of Eq.(\ref{26}) i.e $\displaystyle{\frac{\partial S}{\partial x}}$ and that implies a probability 
evolution equation
\bea
\label{29}
\frac{\partial P}{\partial t}=D\frac{\partial^2 P}{\partial^2 x}
\eea
If $x$ is $x_0$ at $t=t_0$, then the probability $P(x,t,|x_0,t_0)$ at a subsequent time is easily seen to be
\bea
\label{30}
 P(x,t,|x_0,t_0)=\frac{1}{\sqrt{2 \pi D (t-t_0)}}\exp(-\frac{(x-x_0)^2}{2D(t-t_0)})
\eea
We can derive the mean square displacement as 
\ber
\label{31}
<(\triangle x)^2> &=& <(x-x_0)^2>= \int_{-\infty}^{\infty} dx \frac{(x-x_0)^2}{\sqrt{2 \pi D(t-t_0)}} 
\exp(-\frac{(x-x_0)^2}{2D(t-t_0)}) \nn \\
&=&2D(t-t_0)=2D\Delta \tau
\eer
which brings us back to Eq.(\ref{10}) once more.

This way of formulating the problem allows us to study an example of crossover behaviour in
a straight forward fashion. Instead of allowing the randomly moving particle all of space  to wander about ,
we restrict the motion in a region of length L (we will stick to one dimensional case, the generalisation is
obvious). The particle reflects from the boundary walls placed at $x=0$ and $x=L$. It is clear that the 
mean square displacement has to be of {\it O}($L^2$) independent of time as time goes on increasing indefinitely.
On the other hand Einstein's relation tells us that for $L \rightarrow \infty$ the mean square displacement 
has to be order $\tau$. Thus,
\ber 
\label{32}
(\triangle x)^2 &\approx& L^2  \quad \textrm{for} \quad \tau \rightarrow \infty \quad \textrm{at finite L.} \nn  \\
\nn \\
 \textrm{and} \nn \\  
\nn \\
(\triangle x)^2 &\approx& \tau \quad \textrm{for}\quad L \rightarrow \infty 
\eer
This is a classic crossover problem, the crossover being determined by the scales $L$ and $\tau$. 
For $L^2 >> \tau$ it is the usual Einstein relation, while for $ L^2 << \tau$ it is a size limited 
answer. The crossover function which interpolates between the two limits can be constructed by solving 
Eq.(\ref{29}) in a finite geometry. 
\begin{figure}
\centering
\includegraphics[width=5cm,height=7cm]{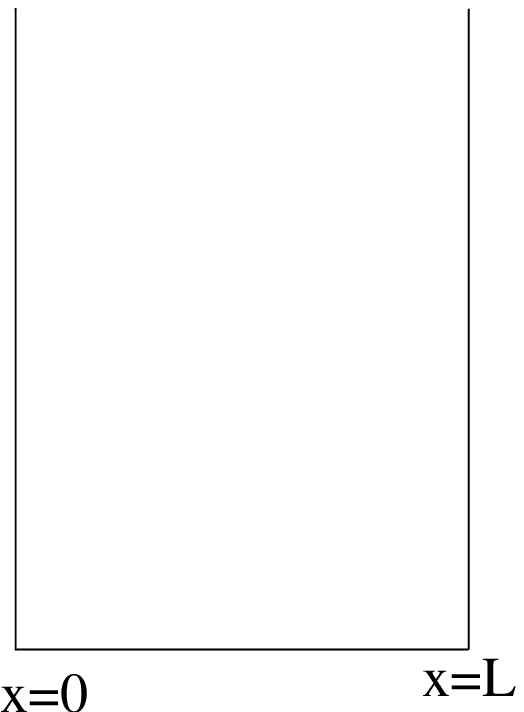}
\caption{}
\end{figure}
The normalized version of Eq.(\ref{29}) with the boundary condition that 
$\displaystyle{\frac{\partial P}{\partial x}=0}$ (reflects from the wall, no current) at $x=0$ 
and $x=L$, can be written as
\bea
\label{33}
P(x,t|0,0)=\frac{1}{L}+\frac{2}{L}\sum_{n=1}^{\infty} \cos(\frac{n \pi x}{L})\exp(\frac{-n^2 \pi^2 D t}{L^2})
\eea
The mean square displacement is readily found to be
\ber
\label{34}
<(\triangle x)^2>&=&\frac{L^2}{3}+\frac{4L^2}{\pi^2}\sum_{n=1}^{\infty}
 \frac{(-1)^n}{n^2} \exp(\frac{-n^2 \pi^2 D t}{L^2})  \\
\label{35}
 &=& \frac{L^2}{3}-\frac{4L^2}{\pi^2}\sum_{\textrm{\it n odd}} \frac{1}{n^2} \exp(\frac{-n^2 \pi^2 D t}{L^2})
+\frac{4L^2}{\pi^2} \sum_{\textrm{\it n even}} \frac{1}{n^2} \exp(\frac{n^2 \pi^2 Dt}{l^2}) \nn \\
\eer
It is straight forward to check that 
\ber
\label{36}
<(\triangle x)^2> &\rightarrow& \infty \quad \textrm{for} \quad t \rightarrow \infty \nn \\
\nn \\
\textrm{and} \nn \\
\nn \\
<(\triangle x)^2> &\rightarrow& 2Dt \quad \textrm{for} \quad L \rightarrow \infty 
\eer
To find the crossover function, we need to express Eq.(\ref{35}) in a closed form. This can be done
by using the Euler-Maclaurin sum formula for the two sums shown in Eq.(\ref{35}). The sum formula is given by
\ber
\label{37}
\int_{a}^{b}f(x) dx=h [\frac{1}{2}f(a)+f(a+h)+f(a+2h)+\cdots + f(b-h)+\frac{1}{2}f(b)] \nn \\
-\frac{1}{2!}B_2 h^2 f'(x)\biggr |_a^b -\frac{1}{4!} B_4 h^4 f'''(x)\biggr |_a^b \nn \\
\eer
where $B_2$ and $B_4$ are the Bernoulli numbers with
\bea
\label{38}
 B_2=\frac{1}{6} \quad \textrm{and} \quad B_4=-\frac{1}{30}
\eea
Keeping only $\displaystyle{exp(-\frac{\pi^2 Dt}{L^2})}$ from among the different exponential decays and working to 
{\it O}($\frac{t}{L^2}$) the coefficient of $\displaystyle{exp(-\frac{\pi^2 Dt}{L^2})}$, the crossover function can be 
written as
\ber
\label{39}
<(\triangle x)^2> = \frac{L^2}{3}-\frac{2L^2}{\pi^2} \int_1^2 \frac{1}{n^2}  \exp(-\frac{n^2 \pi^2 Dt}{L^2})dn  \nn \\
+L^2(\frac{1}{\pi^2}-\frac{1}{3})(1+\frac{D \pi^2 t}{L^2}) \exp(-\frac{D \pi^2 t}{L^2})
\eer
Going for a stronger approximation, we can write the crossover function as
\ber
\label{40}
<(\triangle x)^2> = \frac{L^2}{3}\biggl[1-(1+\frac{D(\pi^2-6)t}{L^2}) \exp(-\frac{\pi^2 Dt}{L^2})\biggr]
\eer
The numerical simulation of the finite size diffusion and it's comparison with our
crossover function are shown in Fig.5.
\begin{figure}
\centering
\includegraphics[width=14cm,height=9cm]{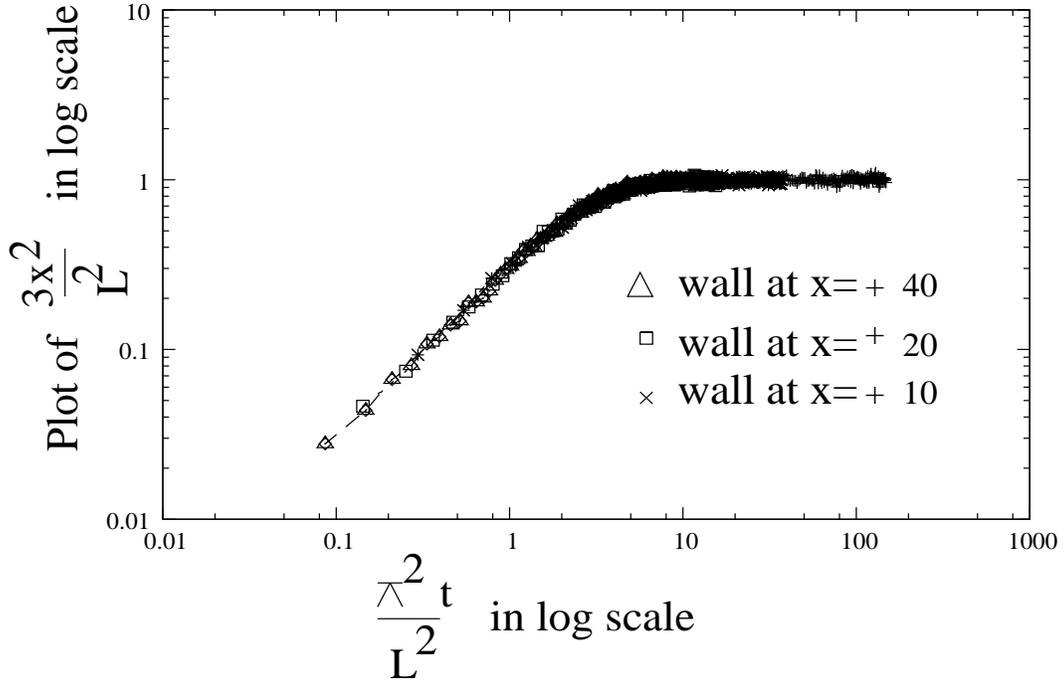}
\caption{Plot of $\frac{3 x^2}{L^2}$ vs $\frac{\pi^2 t}{L^2}$ for different values of L}
\end{figure}

The Langevin equation approach has been of tremendous importance in extending Einstein's 
picture of molecular motion to several systems over the last one hundred years.One of the prime
areas in this respect has been the dynamics near second order phase transitions. For over three decades,
the nineteen sixties,seventies and eighties,critical dynamics had been one of the frontier areas of research
in condensed matter physics. A typical example would be the liquid gas critical point which was 
found by Andrews almost one hundred years ago. In his extensive experiment on carbon dioxide, Andrews found
that at low temperatures as one increases the pressure on a dilute gas, the volume decreased and 
at a particular density the gas condensed to a liquid and over a range of density there were 
coexisting liquid and vapour phases. The pressure had to be increased enormously to cause a ........
of density after the whole gas had liquified. However, about a critical temperature $T_C$, 
the gas could not be liquified by application of pressure. Consequently at $T_C$, the coexisting 
phase diagram disappears and one has a single phase region. The transition at $T_C$ from a one phase system
to a coexisting phases system is a second order transition and is characterised by large scale fluctuations.
As one lowers the temperature towards the critical point, droplets of the liquid phase are formed due to 
the fluctuations and the droplets become larger and larger as the critical point is approached. If the 
characteristic size of the droplet is $\xi$, the $\xi$ becomes infinitely big at $T=T_C$ since the droplets
then become of macroscopic size and at $T_C$ the liquid and vapour phases coexist. This divergence is in
general characterised by an exponent $\nu$ and one writes
\bea
\label{41}
\xi=\xi_0(T-T_C)^{\nu}
\eea
where $\xi_0$ is a system dependent constant and $\nu$ is system independent constant.

We now imagine a thermal gradient applied to this system. The droplets diffuse to carry the heat current
and as the size of the droplets increase the process is expected to become more and more inefficient.
This is what we will be able to demonstrate by using the arguments that we used in the early part of the 
article. A temperature gradient produces a pressure gradient and a force per droplet which can be written
as
\bea
\label{42}
F=-k_B \frac{dT}{dx}
\eea
The viscous forces on the droplet is $\displaystyle{6 \pi \eta \, \xi v}$, which yields
\bea
\label{43}
v=\frac{k_B}{6 \pi \eta \,\xi} \frac{dT}{dx}
\eea
The heat current can be written as 
\bea
\label{44}
j_{\scriptscriptstyle h}= T C_p v
\eea
where $C_p$ is the constant pressure specific heat per unit volume. The heat current by definition
is
\bea
\label{45}
j_{\scriptscriptstyle h}=- \lambda \frac{dT}{dx}
\eea
where $\lambda$ is the thermal conductivity.
From Eq.(\ref{44}) and Eq.(\ref{45}) we find
\bea
\label{46}
\frac{\lambda}{C_p}=\frac{k_{\scriptstyle B} T}{6 \pi \eta \, \xi}
\eea
where $\displaystyle{\frac{\lambda}{C_p}}$ is the coefficient of heat diffusion. As expected
diffusion becomes very small as $T$ approaches $T_C$ because of the divergence of the correlation 
length $\xi$. On the other hand, the specific heat at constant pressure diverges as $T_C$ is approached
and the divergence goes $\xi^2$. Consequently, Eq.(\ref{46}) leads to the remarkable result that the thermal 
conductivity of a fluid diverges near the critical point since
\bea
\label{47}
\lambda=\frac{k_{\scriptstyle B} T C_p}{6 \pi \eta \, \xi} \quad \propto \, \xi
\eea
This spectacular result was first obtained experimentally by Jan Sengers in the early sixties and 
the theoretical result established in the seventies. It is interesting to see how the simple analysis
carried out by Einstein can be invoked to understand a path breaking result found more than fifty 
years later.

We now turn to another aspect of diffusion which has seen significant alterations in the course of last 
one hundred years. If we drop a lump of sugar in a beaker of water and let molecular diffusion cause 
the sugar to dissolve in water, then it is a common experience that it takes a long time for the sugar to 
mix uniformly in the liquid. We can use Einstein's relation to estimate the time. If $L$ is the 
typical dimension of the beaker, then according to Eq.(\ref{10}) the mixing time $\tau_m$ is given by
\ber
\label{48}
\tau_m & = &\frac{L^2}{2D} \, \, \simeq \frac{10^2 \,\mathrm{cm}^2}
{2 \times 10^{-2} \, \mathrm{cm}^2/\mathrm{sec}}  \\
&\simeq& 1.4 \, \mathrm{hrs}\nn
\eer
for a beaker with $L \simeq 10 \,\mathrm{cm}$ and a diffusion coefficient of 
$ 10^{-2} \, \mathrm{cm}^2/\mathrm{sec}$ (typical of sugar in water). This is an awfully long time and 
that is why liquid is stirred in order to make the sugar mix uniformly in a short time. If $v$ is the 
typical velocity, then the mixing is now 
\bea
\label{49}
\tau_m^{'}=\frac{L}{v}
\eea
Clearly the ratio of the two mixing times is given by
\ber
\label{50}
\frac{\tau_m^{'}}{\tau_m}&= &\frac{L}{v}\cdot \frac{2D}{L^2}=\frac{2\nu}{vL}\cdot \frac{D}{\nu} \nn \\
&=& \frac{2S}{Re}
\eer
where $\nu$ is the momentum diffusivity that is $\displaystyle{\nu=\frac{\eta}{\rho}}$ 
(kinematic viscosity)
and $S \,\, \textrm{and} \,\, Re$ are two dimensionless number known as Schmidt number and Reynolds number 
respectively. The Schmidt number $\displaystyle{S=\frac{D}{\nu}}$ is the ratio of the two diffusivities
$D$ and $\nu$ and is a number of order unity. The Reynolds number $\displaystyle{Re=\frac{vL}{\nu}}$ 
requires an analysis of fluid flow.

The dynamics of an incompressible fluid is governed by Navier Stokes equation which reads
\bea
\label{51}
\frac{\partial \vec v}{\partial t} + (\vec v \cdot \vec \nabla)\vec v = - \frac{\vec \nabla P}{\rho}
+\nu \nabla^2 \vec v
\eea
The terms on the right hand side are the forces-a force coming from a pressure 
gradient and a force coming from the viscous drag. On the left hand side, 
we have total acceleration. The first term originates from an explicit 
time dependence of the velocity and the second from the fact that the flow 
brings in and takes out fluid from an elementary volume and thus causes change 
in momentum of the fluid in the volume considered. The second term in the 
left hand side is nonlinear and the source of all the difficulties in 
the solution of Navier-Stokes equation. This is the term which becomes 
important when the velocity is large and causes instability of laminar flows 
and eventually gives rise to turbulence. How important is the term with 
respect to the other terms? The estimate of this term (called the inertial 
force) is $\displaystyle{\frac{v^{2}}{L}}$ where $v$ is typical velocity and $L$ a typical
length. The estimate of the viscous drag is $\displaystyle{\frac{\nu v}{L^{2}}}$ and thus

\ber
\label{52}
\frac{Inertial force}{Viscous drag} &=& \frac{v^{2}}{L} \frac{L^{2}}{\nu v} 
\nn\\
&=& \frac{L v}{\nu} = Re
\eer

Thus the Reynold's number of a flow is an indicator of how important the 
nonlinear term is in comparison to the linear term. Larger the 
Reynold's number, the greater the role of the nonlinear term. We now see
from Eq.(\ref{50}), that the mixing time is going to get smaller as the Reynold's
number is increased and in the limit of $Re \leftarrow \infty $ 
(fully developed turbulence), the mixing time $\tau_{m}$ in the presence of 
stirring seems to go to zero which is an indication of the fact that in 
this limit, Eq.(\ref{10}) needs to be modified. Turbulent diffusion differs very 
strongly from ordinary diffusion and is the mechanism behind the fast 
dissolution of sugar in a stirred beaker of water. The study of diffusive behaviour has 
drawn a lot of attention over the last couple of decades.

The dissolution by stirring certainly requires us to study a forced 
Navier-Stokes equation. This can be written as
\bea
\frac{\partial v_{\alpha}}{\partial t} 
+v_{\beta} \partial_{\beta} v_{\alpha} =-\frac{1}{\rho} \partial_{\alpha}P
+\nu \nabla^2 v_{\alpha} +f_{\alpha} \nn
\eea
where $\, \displaystyle{f_{\alpha}}$ is a random function (specified by 
correlations only) that models the stirring of the fluid. The fluid is 
incompressible which means that $\, \displaystyle{ \vec\nabla\cdot\vec v =0}$.
It is important to note the energy balance of the above equation. The total
energy of the fluid can be written as 
\bea
E=\frac{1}{2} \int v_{\alpha} v_{\alpha} \, d^3 r  \nn
\eea
\bea
\frac{\partial E}{\partial t} = -\int v_{\alpha} v_{beta} 
\partial_{\beta} v_{\alpha}\, d^3 r -\frac{1}{\rho} \int v_{\alpha} 
\partial_{\alpha} P \, d^3 r + \nu \int v_{\alpha} \nabla^2 v_{\alpha}\,d^3 r
+\int f_{\alpha} v_{\alpha}\, d^3 r \nn
\eea
\ber
\int v_{\alpha} \partial_{\alpha} P \, d^3 r &=& 
\int \partial_{\alpha} (Pv_{\alpha})\, d^3 r \nn \\
&=& \int P v_{\alpha} \, dS_{\alpha} =0 \nn
\eer
where we have assumed that the velocity vanishes in distant surfaces and have 
used $\, \displaystyle{\partial_{\alpha} v_{\alpha} = 0}$ in the first step.
In a similar manner
\ber
\int v_{\alpha} v_{\beta} \partial_{\beta} v_{\alpha} \, d^3 r 
&=& \frac{1}{2} \int v_{\beta} (v_{\alpha}v_{\alpha})\, d^3 r \nn \\
&=& \frac{1}{2} \int \partial_{\beta} (v^2 v_{\alpha})\, d^3 r \nn \\
&=& \frac{1}{2} \int v^2 v_{\beta}\, dS_{\beta} \nn \\
&=&0 \nn
\eer
For the viscous dissipation, we have 
\bea
\nu \int v_{\alpha} \nabla^2 v_{\alpha}\,d^3 r = -\nu \int (\partial_{\beta}
v_{\alpha})(\partial_{\beta} v_{\alpha})\, d^3 r \nn
\eea
Thus,
\bea
\frac{\partial E}{\partial t}=-\nu \int (\partial_{\beta} v_{\alpha})^2 \, 
d^3 r + \int f_{\alpha} v_{\alpha}\, d^3 r \nn
\eea
We can maintain a steady state in the system 
$\,\displaystyle{\biggl ( \frac{dE}{dt}=0 \biggr)}$ if we balance the
viscous dissipation (first term on the right hand side) by the second 
term which represents the rate at which energy is fed into the system. In 
this global analysis, the nonlinear term plays no role. Their contribution
is in the carrying of the energy across the different length scales. 
The dissipation term which carries the highest derivatives in the problem is
clearly most relevant at short length scales (dissipation effective at 
molecular scale), while the energy input (no derivatives) is operative 
at the highest length scales. Consequently the energy balance picture
for fully developed turbulence that was arrived at Kolmogorov involves 
injecting energy into the system at large length scales at a constant rate
$\, \displaystyle{\epsilon}$, transferring at from large scales to short 
scales at the constant rate $\, \displaystyle{\epsilon}$ and dissipating it at 
the shortest (molecular) scales by viscous action at the same rate
$\, \displaystyle{\epsilon}$. For this picture to be consistent, the 
highest and shortest length scales need to be well separated. Now, the scale,
$L$,at which energy is injected can be estimated from
\bea
\epsilon = \frac{v^2}{T} \propto \frac{v^3}{L} \nn
\eea 
where $v$ is the typical velocity and $T$ is a typical time scale over which 
the energy containing eddies of size $L$  can turn in response to the drive.
The dissipation scale is $l$ and has to be constructed out of $\epsilon$ and
the kinematic viscosity $\nu$. Dimensional analysis shows that
\bea
l \sim \frac{\nu^3}{l} \nn
\eea
Consequently,
\ber
\frac{L}{l} &\sim& \frac{L}{(\frac{\nu^3}{\epsilon})^{\frac{1}{4}}} \sim
L\cdot \biggl(\frac{v^3}{L}\biggr)^{\frac{1}{4}} \frac{1}{\nu^{\frac{3}{4}}} \nn \\
&=&\biggl(\frac{vL}{\nu}\biggr)^{\frac{3}{4}} = Re \nn
\eer
Thus for high Reynold's number, the scales are extremely well separated.we can see
a particularly relevant feature of turbulent flow from this simple analysis. In the 
limit of $Re$ tending to infinity, the scale $l$ has to be extremely small and if we
try to construct derivative of the velocity field, that derivative will become 
singular, which is exactly what is needed to keep 
$\, \displaystyle{\nu \int d^3 r (\partial_{\alpha} v_{\beta})^2}$ finite in the 
limit of $\, \displaystyle{ \nu \rightarrow 0}$.

We now note that equations of the form 
\bea
\frac{\partial u}{\partial t} = \nabla^2 u \nn
\eea
have solutions which depend on the combination $\, \displaystyle{\frac{r^2}{t}}$.
The solutions can change by a scale factor if $r$ and $t$ are scaled by appropriate
factors. If $r$ is scaled by $\lambda$ and $t$ is scaled by $\lambda^2$, then 
the solutions of $u$ can change only by a power of $\lambda$. Such solutions are 
known as scaling solutions. It should be apparent that Einstein's result for
Brownian motion (Eq.(\ref{10}) is a manifestation of this scaling solution. If 
the diffusion law needs to be changed for the stirred fluid, then there has to be
a different scaling solution. Accordingly, we explore the scaling solution of 
\bea
\frac{\partial v_{\alpha}}{\partial t} + v_{\beta} \partial_{\beta} v_{\alpha} =
\nu \nabla^2 v_{\alpha}  \nn 
\eea
We have dropped the pressure term because taking a divergence of Eq.(\ref{51}),
we note that 
\bea
\frac{1}{\rho} \nabla^2 P = \partial_{\alpha} v_{\beta} \partial_{\beta} v_{\alpha}\nn
\eea
and thus the term $\,\displaystyle{\partial_{\alpha} \frac{P}{\rho}}$ can be written
as $\,\displaystyle{\partial_{\alpha}\biggl[ 
\frac{\partial_{\alpha} v_{\beta} \partial_{\beta} v_{\alpha}}{\nabla^2}\biggr]}$
and the dimension of this term is identical to that of 
$\,\displaystyle{v_{\beta}\partial_{\beta} v_{\alpha}}$.
Consequently, the pressure term cannot modify a dimensional analysis.

We carry out the following scaling
\ber
\vec x & =&\lambda \vec x' \nn \\
t &=& \lambda^z t' \nn \\
\vec v &=& \lambda^{\frac{\mu}{3}} \vec v' \nn
\eer
In the primed variables, we have 
\bea
\lambda^{(-z+\frac{\mu}{3})} \frac{\partial v'_{\alpha}}{\partial t'}
+\lambda^{(\frac{2\mu}{3}-1)} v'_{\beta}\partial'_{\beta} v'_{\alpha}
=\nu \lambda^{(\frac{\mu}{3}-2)} \nabla'^{2} v'_{\alpha} \nn
\eea
or
\bea
\frac{\partial v'_{\alpha}}{\partial t'}
+\lambda^{(\frac{\mu}{3}+z-1)} v'_{\beta}\partial'_{\beta} v'_{\alpha}
=\nu \lambda^{(z-2)} \nabla'^{2} v'_{\alpha} \nn
\eea
If we impose 
\bea
z=1-\frac{\mu}{3} \nn
\eea
then provided $\nu$ is scale dependent and becomes
\bea
\nu = \nu' \lambda^{(2-z)}  \nn
\eea
We have the same Navier-Stokes equation in primed variables
\bea
\frac{\partial v'_{\alpha}}{\partial t'}+v'_{\beta}\partial'_{\beta} v'_{\alpha}
=\nu' \nabla'^{2} v'_{\alpha} \nn
\eea
The non-linear term has had two drastic effects \\
\\
(a) z is no longer 2 and hence Eq.(\ref{10}) will have to be modified. \\
\\
(b) the viscosity has become scale dependent-it is no longer the molecular
viscosity. We have generated a turbulence induced scale dependent viscosity.
\\

The scaling exponent $z$ is however unknown yet. To explore that further we 
explore the effect of the scale transformation on the energy flow per unit time 
$\, \displaystyle{\epsilon}$. We find 
\bea
\epsilon'= \frac{v'^3}{L'^3}=\frac{\lambda^{\mu} v^3}{\lambda L}
=\lambda^{\mu-1} \epsilon \nn
\eea
Since the rate of energy flow is constant, we require 
$\, \displaystyle{\epsilon=\epsilon'}$ and that fixes $\,\displaystyle{\mu=1}$,
leading $\,\displaystyle{z=\frac{2}{3}}$. We now explore the effect of this on
the concentration evolution. The scale transformation takes 
$\, \displaystyle{\frac{\partial c}{\partial t}+(\vec v \cdot \vec \nabla)c 
=D \nabla^2 c}$ to,
\bea
\lambda^{-z} \frac{\partial c'}{\partial t'}+\lambda^{(\frac{\mu}{3}-1)}
(\vec v' \cdot \vec \nabla')c'=D\lambda^{-2} \nabla'^2 c' \nn
\eea
or
\bea
\frac{\partial c'}{\partial t'}+\lambda^{(\frac{\mu}{3}+z-1)}
(\vec v' \cdot \vec \nabla')c'=D' \nabla'^2 c' \nn
\eea
where
\bea
D'=\lambda^{2-z} D \nn
\eea
The advective terms in the diffusion equation has thus induced a scale dependent
diffusion coefficient, where $D$ has to scale as $L^{2-z}$. In Eq.(\ref{10}),
we now need the structure
\bea
(\triangle x)^2 = D_0 \biggl( \frac{L}{l} \biggr)^{2-z} t \nn
\eea
where $D_0$ has the dimension of a diffusion constant and $l_0$ is a 
suitable length scale. In terms of $t$, $\, \displaystyle{L \sim t^{\frac{1}{z}}}$
and thus,
\bea
(\triangle x)^2 \, \propto \,D_0 \,\frac{t^{\frac{2}{z}-1}}{l_0^{2-z}} \,t \, 
\propto \,t^{\frac{2}{z}}\, = \,t^3 \nn
\eea
where we have used $z=\frac{2}{3}$ in the last step. Consequently, for turbulent
diffusion
\bea
(\triangle x)^2 \, \propto\,t^3 \nn
\eea
 which replaces Eq.(\ref{10}) of Einstein and ensures in a stirred fluid, 
the mixing takes place much faster than it would otherwise.

We end this article by considering a particular development
which has occurred over the last decade.It brought to light
a feature of the diffusion process which had not been noted before.
The diffusion equation for a concentration $n(x,t)$ is a conservation
law for the total concentration. In the region of thickness $\Delta x$
in fig.1, the amount of diffusing material is $n(x,t)A\Delta x$, where
$A$ is the cross sectional area. In time $\Delta t$, the change in amount 
of material is $\,\displaystyle{\frac{\partial n(x,t)}{\partial t} A 
\Delta x \Delta t}$.
This change can only be brought about the solute current (this is the 
conservation law) and is the difference in the current at the two surfaces
bounding the region of width $\Delta x$. The net amount of material
flowing into the region in time $\Delta t$ because of the difference 
in current at the two surfaces is $- \frac{ dj}{dx} \Delta x A \Delta t$.
The conservation law then reads
\bea 
\label{53}
\frac{\partial n}{\partial t} 
= -\frac{dj}{dx} = D\frac{\partial^{2}n}{\partial x^{2}}
\eea

on using Eq.(\ref{6}). We can without any loss of generality rescale $'x'$ 
and $'t'$ to set $D=1$ and we can write the diffusion equation as
\bea
\label{54}
\frac{\partial n}{\partial t}=\frac{\partial^{2}n}{\partial x^{2}}
\eea

This solution is well known and can be written as
\bea
\label{55}
n(x,t)=\int G(x-x',t) n(x',0)dx'
\eea

where
\bea
\label{56}
G(x-x^{\prime},t)=\frac{1}{(4 \pi t)^{1/2}}\exp{-(x-x^{\prime})^{2}/4t}
\eea 

Clearly, the solution is characterised by a single length scale
which grows as $t^{1/2}$. What was discovered by Majumdar et al 
about ten years ago is that there is another non-trivial exponent 
associated with this apparently solved problem.

To understand the origin of this new exponent, we consider a class of initial 
conditions where $n(x,0)$ is a Gaussian random variable with zero mean. 
The question asked is : what is the probability $p_0(t)$ that the field $n(x,t)$
at a particular point $x$ has not changed its sign till time t. For large
time, this probability has to decay to zero and this long time decay is 
charactrized by the law 
\bea
\label{57}
p_0(t) \propto t^{-\theta}
\eea
The exponent $\theta$ is the new non-trivial exponent that we talked about
and is called the persistence exponent. It also turns out that 
$\displaystyle{p_n(t_1,t_2)}$ - the probability that the field 
changes sign $n$ times between $\displaystyle{t_1}$ and 
$t_2$ ($t_1$ $>$ $t_2$) - is given by 
\bea
\label{58}
p_n(t_1,t_2) \sim c_n \biggl[ \ln \frac{t_1}{t_2}\biggr]^n 
\biggl(\frac{t_1}{t_2}\biggr)^{-\theta}
\eea
The primary feature of the calculation is the point that the Gaussian 
process $n(x,t)$ is a Gaussian stationary process in terms of the new variable 
$T=\ln t$. This is followed by the central assumption that the interval between
two successive zeros of $n(x,t)$ can be treated as independent.

The distribution of $n(x,0)$ is taken to be Gaussian and white, i.e
the two point correlation is
\bea
\label{59}
< n(x,0) n(x',0)> = \delta (x-x')
\eea
We can now write down the correlator as 
\ber
\label{60}
< n(x,0) n(x',0)> &=& <\int \, dx' \, G(x-x',t_1) \phi(x',0) 
\int\, dx''\, G(x-x'',t_2) \phi(x'',0)> \nn \\
&=& \iint\, dx'\, dx''\, G(x-x',t_1)G(x-x'',t_2)<\phi(x',0)\phi(x'',0)> \nn \\
&=& \int \,dx'\, G(x-x',t_1)G(x-x',t_2)
\eer
where we have made the use of Eq.(\ref{59}) in arriving at the last step. Using the Green function given in Eq.(\ref{56}),
\ber
\label{61}
<n(x,t_1)n(x',t_2)> &=& \frac{1}{4 \pi \sqrt{t_1 t_2}} 
\int_{-\infty}^{\infty} \, dx' \, 
\exp\biggl(-\frac{(x-x')^2}{4 t_1 t_2} (t_1 +t_2)\biggr) \nn\\
&=& \frac{1}{2 \sqrt{\pi} } \cdot \frac{1}{(t_1 +t_2)^{\frac{1}{2}}}
\eer
We immediately see that
\bea
<n(x,t_1)^2>= \frac{1}{2 \sqrt{2 \pi}\, t_1^{\frac{1}{2}}} \nn
\eea
If we now define a variable $\,\displaystyle{X(x,t)=\frac{n(x,t)}
{\sqrt{n(x,t)}}
=2^{\frac{1}{2}} \,(2 \pi)^{\frac{1}{4}}\, t^{\frac{1}{4}}\, n(x,t) }$,
then the correlation function 
\ber
\label{62}
a(t_1, t_2) &=& <X(t_1)X(t_2)> = \frac{ 2 (2 \pi)^{\frac{1}{2}} 
(t_1 t_2)^{\frac{1}{4}}}{2 \sqrt{\pi} (t_1 + t_2)^{\frac{1}{2}}} \nn \\
&=& \biggl[ \frac{4 t_1 t_2}{(t_1 +t_2)^2} \biggr]^{\frac{1}{4}}
\eer
This is not yet stationary process. To construct a stationary process we 
need to use the new time variable $T= \ln t$. Now
\ber
\label{63}
a(T_1,T_2)&=&\biggl[ \frac{4 \exp(T_1+T_2)}
{\displaystyle{(e^{T_1}+e^{T_2})^2}} \biggr] \nn \\
&=& \biggl[ \frac{4 \displaystyle{e^{(T_2 - T_1)}}}
{(1-\displaystyle{e^{(T_2-T_1)}})} \biggr] \nn \\
&=&\biggl[ \mathrm{sech}\lbrace \frac{1}{2}(T_2-T_1) \rbrace \biggr]
\eer
Thus one has arrived at a Gaussian stationary process. the anticipated 
form of $\displaystyle{p_0(t_1,t_2)}$ is $\,\displaystyle{p_0(t_1,t_2)
\sim \biggl ( \frac{t_1}{t_2}\biggr )^{\theta} \sim e^{-\theta (T_2-T_1)}}$.
In the new time variable the calculation of $\theta$ is the calculation 
of decay rate. In the process of calculation, one needs to make a very 
important assumption - the interval between zeros of $X(T)$ are statistically
independent.

The correlation function $a(T_2 - T_1)$ will not give us the information 
regarding the zeros of $X(T)$. For that we need to know teh correlation 
of the sign of $X(T)$. Accordingly we define a function 
$\, \displaystyle{\sigma(T)=sgn X(T)}$ and consider the correlation
\bea
\label{64}
A(T)= <\sigma(0)\, \sigma(T)>
\eea
The correlation function is determined entirely by the distribution 
$P(T)$ of the intervals between zeros. The trick is to find the $P(T)$
from $A(T)$ and then $p_0(T)$ from $P(T)$. The first step is knowing $A(T)$
and this is possible for a Gaussian stationary process , because 
for such a process 
\bea
\label{65}
A(T)=\frac{2}{\pi} \sin^{-1} a(T)
\eea
Since $a(T)$ has been calculated in Eq.(\ref{63}) we know $A(T)$. Now 
if $p_n(T)$ is the probability of there being $n$ zeros between 0 and $T$,
then clearly
\bea
\label{66}
A(T)=\sum_{n=0}^{\infty} (-1)^{n} p_n(T)
\eea
Looking at the interval between 0 and $T$, if there is a zero at $T_1$, 
then to find $p_1(T)$, the probability of there being just one zero 
between 0 and $T$ then we nedd to introduce $Q(T)$, the probability
that an interval to the right or left of a zero contains no further 
zeros. Clearly
\bea
\label{67}
p_1(T)=\frac{1}{<T>} \int_0^{T} \, dT_1 \, Q(T_1) Q(T-T_1)
\eea
where $<T>$ is the mean interval size. Now $P(T)$ is the probability
of finding an interval of length $T$ between two successive zeros.
So, if we are interested in the probability of there being two zeros
between 0 and $T_1$ (the zeros being at $T_1$ and $T_2$ ), then
\bea
\label{68}
p_2(T)=\frac{1}{<T>} \int_0^{T} \, dT_1 \int_{T}^{T_1} \, dT_2 \,
Q(T_1) P(T_2-T_1) Q(T-T_2)
\eea
if the intervals are all independent of each other. This is the 
{\it "Independent Interval Approximation"} and generalising to the 
case of $n$ zeros,
\ber
\label{69}
p_n(T)=\frac{1}{<T>} \int_0^{T}dT_1 \int_{T_1}^{T_2}dT_2 \dots
\int_{T_{n-1}}^{T}dT_n  Q(T_1)P(T_2-T-1)P(T_3-T_2)\dots \nn \\
\dots P(T_n-T_{n-1})Q(T-T_n) \nn \\
\textrm{for n} \, \ge 1 \nn \\
\eer
The probability that an interval to the left or right of a zero
does not contain a zero is unity if the interval is of unit length
and hence
\bea
Q(T)=1-\int_0^{T} P(t) dt \nn
\eea
which leads to 
\bea
\label{70}
Q'(T)=-P(T)
\eea
Using Laplace transforms, if $\displaystyle{\widetilde{P}(s)}$ and 
$\displaystyle{\widetilde{Q}(s)}$ are the transforms of $P(T)$ and 
$Q(T)$, then
\bea
\label{71}
\widetilde{P}(s)=1-s\widetilde{Q}(s)
\eea
and from Eq.(\ref{69})
\bea
\label{72}
\tilde{p}_n (s) =\frac{1}{<T>}[\widetilde{Q}(s)]^2 [\widetilde{P}(s)]^{n-1}
\eea
Eliminating $\displaystyle{\widetilde{Q}(s)}$ by using Eq.(\ref{71})
we get
\bea
\label{73}
\tilde{p}_n (s)= \frac{1}{<T> s^2} [1-\widetilde{P} (s)]^2
[\widetilde{P} (s)]^{n-1} \quad \quad \, \textrm{for n} \, \geq 1
\eea
For $n=0$, we use $\,\displaystyle{ \sum_{n=0}^{\infty} p_n(t) =1}$,
which implies $\, \displaystyle{ \sum_{n=0}^{\infty} p_n(s)= \frac{1}{s}}$
to write
\bea
\label{74}
\tilde{p}_0 (s)= \frac{1}{s} -\sum_{n=1}^{\infty} \tilde{p}_n (s)
= \frac{1}{s} -\frac{1}{<T> s^2} [1-\widetilde{P} (s)]
\eea
The Laplace transform of Eq.(\ref{66}) gives 
$\, \displaystyle{ \widetilde{A} (s) = \sum_{n=0}^{\infty} 
(-1)^n \tilde{p}_n (s)} $. We use Eq.(\ref{73}) and Eq.(\ref{74}) to perform
the sum and finally obtain
\bea
\label{75}
\widetilde{P} (s) =\frac{[2-F(s)]}{F(s)}
\eea

\bea
\label{76}
F(s)=1+\frac{<T>}{2}\cdot s [1-s\widetilde{A} (s)]
\eea
The sequence now is as follows: From Eq.(\ref{76}), $F(s)$ is obtained in 
terms of $\widetilde{A}(s)$ and from Eq.(\ref{75}), $\widetilde{P}(s)$ is
obtained in terms of $\widetilde{A}(s)$. We not that $A(T)$ has been 
calculated in Eq.(\ref{65}) and Eq.(\ref{63}). Once $\widetilde{P}(s)$
is known $\tilde{p}_0 (s)$ and $\tilde{p}_n(s) \, \, \, [ n \geq 1]$ 
are known from Eq.(\ref{74}) and Eq.(\ref{73}). The inverse Laplace 
transform will yield $ \, \tilde{p}_n(T)$.

It is however simpler to do the following. We note that 
$\, \displaystyle{p_0(T) \sim e^{-\theta} }$ and hence $\tilde{p}_0 (s)$
has a simple pole at $ \,\displaystyle{s= -\theta}$. Looking at the 
structure of Eq.(\ref{74}) and Eq.(\ref{75}), we note that a simple 
pole of $p_0(s)$ is a simple zero of $F(s)$. For the 
$\displaystyle{ \widetilde{A} (s)}$ corresponding to Eq.(\ref{65}),
\bea
\label{77}
F(s)=1+\sqrt{2} \pi s \biggl \lbrace 1-\frac{2s}{\pi} 
\int_0^{\infty} dT \,\displaystyle{e^{-sT} 
sin^{-1}[sech \frac{T}{2}]^\frac{1}{2}} \biggr \rbrace
\eea
This $F(s)$ has only one zero and numerically it is found at $s=-0.1203$
giving
\bea
\theta = 0.1203 \nn
\eea
A numerical simulation of the diffusion equation yields 
$\, \displaystyle{\theta =0.1207}$. Thus, we see how a new nontrivial 
and totally unexpected feature of the diffusion equation has been 
found almost a hundred years after the equation was first used.

\end{document}